\def\year{2021}\relax
\documentclass[letterpaper]{article} % DO NOT CHANGE THIS
\usepackage{aaai21}  % DO NOT CHANGE THIS
\usepackage{times}  % DO NOT CHANGE THIS
\usepackage{helvet} % DO NOT CHANGE THIS
\usepackage{courier}  % DO NOT CHANGE THIS
\usepackage[hyphens]{url}  % DO NOT CHANGE THIS
\usepackage{graphicx} % DO NOT CHANGE THIS
\usepackage{natbib}  % DO NOT CHANGE THIS AND DO NOT ADD ANY OPTIONS TO IT
\usepackage{caption} % DO NOT CHANGE THIS AND DO NOT ADD ANY OPTIONS TO IT
\usepackage[utf8]{inputenc}
\usepackage[english]{babel}

\usepackage[utf8]{inputenc} % allow utf-8 input
\usepackage{url}            % simple URL typesetting
\usepackage{booktabs}       % professional-quality tables
\usepackage{amsfonts}       % blackboard math symbols
\usepackage{amsthm}
\usepackage{nicefrac}       % compact symbols for 1/2, etc.
\usepackage{microtype}      % microtypography
\usepackage{adjustbox}
\usepackage{bbm}            % indicator function
\usepackage{mathtools}
\usepackage{multirow}
\usepackage{enumitem}
\usepackage{attrib}

\newcommand{\ignore}[1]{}
\newcommand{\nop}[1]{}
\newcommand*{\eg}{{\em e.g.}}

\newcommand*{\ie}{{\em i.e.}}

\usepackage{pgfplots,tikz}
\usetikzlibrary{pgfplots.groupplots}
\usetikzlibrary{shapes.geometric}
\usetikzlibrary{positioning,fit,shapes.geometric,backgrounds, calc}
\tikzset{textnode/.style={inner sep=0pt,outer sep=0,execute at begin node={\strut}}}
\tikzstyle{state} = [textnode,circle, draw, inner sep=0pt, outer sep=0]
\tikzstyle{label} = [rectangle, inner sep=0pt,outer sep=0,execute at begin node={\strut}, font=\small]  % text nodes 
\tikzstyle{textbox} = [rectangle, draw=none, fill=none, outer sep=2pt, inner sep=0pt]
\tikzstyle{edge} = [shorten >=0.75pt, auto, line width=0.2pt]
\usetikzlibrary{calc}
\usetikzlibrary{arrows,decorations.markings}
\usetikzlibrary{shapes.arrows}
\usetikzlibrary{patterns}
\usetikzlibrary{spy}
\selectcolormodel{cmyk}
\pgfplotsset{compat=1.15}

\definecolor{burnt_orange}{HTML}{a72207}
\definecolor{burnt_blue}{HTML}{3f4f73}
\colorlet{background_grey}{white!70.9019607843137!black}

\usetikzlibrary{pgfplots.groupplots}
\usetikzlibrary{shapes.geometric}
\usetikzlibrary{positioning,fit,shapes.geometric,backgrounds, calc}
\usetikzlibrary{arrows,decorations.markings}
\usetikzlibrary{shapes.arrows}
\usetikzlibrary{patterns}
\tikzset{textnode/.style={inner sep=0pt,outer sep=0,execute at begin node={\strut}}}
\tikzstyle{state} = [textnode,circle, draw, inner sep=0pt, outer sep=0]
\usepgfplotslibrary{groupplots}

\selectcolormodel{cmyk}

\pgfplotsset{every axis/.append style={
                    xlabel={$x$},          % default put x on x-axis
                    ylabel={$y$},          % default put y on y-axis
                    tick label style={font=\scriptsize\sffamily},
                    title style = {font=\scriptsize\sffamily},
                    ylabel near ticks,
                    y label style={font=\sffamily\scriptsize},
                    xlabel near ticks,
                    x label style={font=\sffamily\scriptsize},
                    legend cell align={left},
                    legend style={draw=none, font=\sffamily\tiny},
                    },
                    legend image code/.code={
                    \draw[mark repeat=2,mark phase=2]
                        plot coordinates {
                        (0cm,0cm)
                        (0.15cm,0cm)        %% default is (0.3cm,0cm)
                        (0.3cm,0cm)         %% default is (0.6cm,0cm)
                        };%
                    }
                    }
\pgfplotsset{compat=newest}  
\usepgfplotslibrary{colorbrewer}   
\pgfplotsset{cycle list/Dark2}

\newcommand{\todo}[1]{\textcolor{red}{TODO: #1}}

\newcommand{\newstuff}[1]{#1}

\title{Subreddit Links Drive Community Creation and User Engagement on Reddit} %When Links to Nowhere Spawn Communities}

\author{
    Rachel Krohn,\textsuperscript{\rm 1}
    Tim Weninger \textsuperscript{\rm 2}
    \\
}
\affiliations{
    \textsuperscript{\rm 1}Rose-Hulman Institute of Technology,\\
    \textsuperscript{\rm 2}University of Notre Dame\\
    krohn@rose-hulman.edu,
    tweninger@nd.edu
}

\begin{document}

\maketitle

\begin{abstract}
On Reddit, individual subreddits are used to organize content and connect users. One mode of interaction is the \textit{subreddit link}, which occurs when a user makes a direct reference to a subreddit in another community. Based on the ubiquity of these references, we have undertaken a study on subreddit links on Reddit, with the goal of understanding their impact on both the referenced subreddit, and on the subreddit landscape as a whole. By way of an extensive observational study along with several natural experiments using the entire history of Reddit, we were able to determine that (1) subreddit links are a significant driver of new suberddit creation; (2) subreddit links (2a) substantially drive activity in the referenced subreddit, and (2b) are frequently created in response to high levels of activity in the referenced subreddit; and (3) the graph of subreddit links has become less dense and more treelike over time. \newstuff{We conclude with a discussion of how these results confirm, add to, and in some cases conflict with existing theories on information-seeking behavior and self-organizing behavior in online social systems.}
\end{abstract}

\iffalse
\todo{"so what?" i.e., what can/should sociologists, users/community maintainers on reddit, etc., take away from these findings and this particular research question? How does analyzing subreddit links contribute to our broader understanding of online discourse, community formation, and reddit?}

\todo{relate the reddit analysis to broader questions surrounding user behaviour, community formation etc}

\todo{contextualize findings in existing work}

\todo{authors illustrate, via their graph analysis, longitudinal trends in how reddit has evolved. How would this reflect, confound, and add nuance to existing findings on community development, intercommunity interactions, and platform dynamics?}

\todo{more concretely connect/contrast to existing work - do a better job of commenting on how this work relates to this existing research and past findings}

\todo{My biggest concern was that the research is not well motivated. It is not clear why studying this is important, what will it accomplish and how does it improve our understanding. While reporting all statistical texts and method details is helpful, the authors could have spent a bit more time discussing the impact of this work.}

\todo{reflect upon how this research is applicable beyond this paper}
\fi

\section{Introduction}

\begin{quote}
\textit{All that is gold does not glitter,\\
Not all those who wander are lost;
}
\attrib{J.R.R. Tolkien, {\em Fellowship of the Ring}}
\end{quote}

\noindent \newstuff{Users of online social systems tend to group into communities based on common interests, a specific topic, or a singular purpose. These communities take on unique characteristics based on their form or function. Depending on the platform, these communities can be well-defined or completely informal. For example, Reddit is particularly community-focused: all content is organized into subreddits and each community is populated by users for the purpose of sharing and discussing relevant content. The same is true in many online social systems like StackExchange, Tumblr, and others. %On the other hand, the Twitter network consists of users that may follow or retweet other users, but communities are not rigidly defined. Instead, online social communities on Twitter are more akin to the organizational dynamics of hashtags~\cite{jackson2018girlslikeus}, but that comparison is limited.
}

\newstuff{In the present work, we focus on explicit inter-community links, \ie, direct, clickable, hyperlinks from one community to another. Hyperlinks like these are the defining characteristic of the Web. Findings of information navigation~\cite{farzan2018social} and human wayfinding~\cite{west2012human} on the Web are largely based on the constructionist interpretation of \textit{information behavior theory}~\cite{fisher2005theories} and have given way to well-defined models of human behavior in these linked systems. These theories and models suggest that hyperlinks are a primary driver of user navigation and attention, and are evidence of user interest~\cite{gao2021examining}.%, but they do not make predictions about the creation of new information nor the formation of communities in linked social systems. 
}

\begin{figure}[t]
    \centering
    \includegraphics[width=0.9\linewidth]{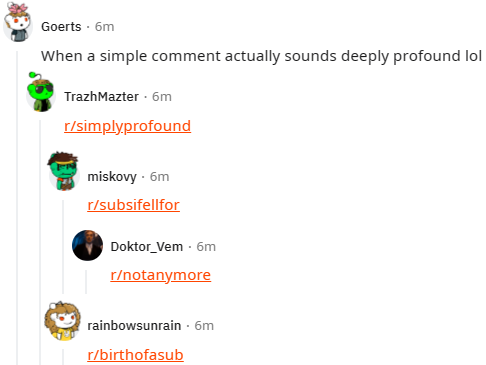}
    \caption{Example of \textit{subreddit links} in a Reddit comment thread. Each link points to a subreddit as a whole, and not to a single post or comment within that subreddit. These links are not only used to promote communities, but are also used as social commentary or jokes. In this case, r/simplyprofound did not exist at the time the link was made, prompting the r/subsifellfor reply. However, the subreddit r/simplyprofound was created as a result of this link, leading to the r/notanymore reply. Additionally, r/birthofasub is referenced, since this community chronicles the creation of new subreddits. \nop{source: https://www.reddit.com/r/nextfuckinglevel/comments/lzn64t/architect_builds_a_3story_dollhouse_out_of/}}
    \label{fig:link_example}
\end{figure}

\newstuff{Hyperlinks are typically assumed to be intentional and reflective of some organizational structure -- a structure typically decided by the organization and implemented by a Web administrator~\cite{weninger2013parallel}. But what happens when information structures and communities are defined by the users themselves? Existing models and theoretical frameworks have surprisingly little to say on the subject. Empirical studies on Wikipedia and Twitter show that users use links (wikilinks and clickable hashtags resp.) to self-organize and drive user attention to content~\cite{bairi2015evolution}. But does the same hold true for social communities? Do online links drive the formation and growth of online social groups? }

Reddit is a natural environment in which to consider these questions because of the well-defined subreddit structure and because inter-community references, illustrated in Fig.\ref{fig:link_example}, are commonplace on Reddit. We call these inter-community references \textit{subreddit links} and, unlike links to content or users, these subreddit links reference an entire community -- \textit{not} a single post or comment. Therefore, users who create these subreddit links are consciously deciding to reference the community as a whole, and not one particular discussion or user within the community. Given that all Reddit content is hosted within some community, we can also say that each subreddit link originates in a particular community. In this way, subreddit links may be viewed as directed connections from one community to another. 

\newstuff{Most subreddit links reference an existing subreddit, encouraging users to explore or join established communities.} In other cases, links are made to subreddits that do not exist (often as a joke), and occasionally these links spur the creation of a new community. As a result subreddit links have the power to alter the community landscape of Reddit. The ability of these links to drive online community development and change has not, until now, been carefully studied.

% stole 2 paragraphs from here, and moved them to the next section

%hi rachel - do you want to move our meeting to now?

Instead, most work has focused on user-level behavior analysis~\cite{tan2015all} or on individual communities and hashtag campaigns, \eg, \cite{jackson2018girlslikeus}. Interaction between community entities is less understood. What role do these inter-community links serve in online communities? And how do these inter-community links affect the evolution and dynamics of the online social system at large?

\newstuff{Answers to these questions are critical to our understanding of social organization in online spaces and can be used to inform design decisions of future online social systems. Only after we understand the dynamics by which communities are created and grow can we begin to define user roles and other behavioral dynamics~\cite{yang2019seekers}.} 

\subsection{Research Questions}

The overarching goal of the current work is to better understand these subreddit links, where they come from, where they lead, and the effect that these links have on the evolution of the community-structure as a whole. Specifically, we ask:

\setlist[enumerate]{wide=0pt, leftmargin=0pt, labelwidth=15pt, align=left}

\begin{enumerate}
    \item[\textbf{RQ1}] Do subreddit links spur the creation of new communities?
    \item[\textbf{RQ2}] Does a subreddit link impact the activity level of the referenced community?
    \item[\textbf{RQ3}] From the perspective of these subreddit links, how has the landscape of subreddit interactions changed over time? What does this tell us about the evolution of Reddit?
\end{enumerate}

\newstuff{We show that subreddit links have substantive impact on both activity levels within existing communities, and on community generation. Users have adopted this mechanism of community creation, promotion, and navigation, thereby imbuing subreddit links with the power to influence the overall social landscape. These community links therefore represent another avenue of innovation on the social Web, one that seeks to better organize the wealth of content by allowing users themselves, instead of ``the algorithm'', to drive engagement, create communities, and organize content.}

\section{Social Media Communities}   % formerly Background

% \todo{cite meme ecology? other recent papers?}

%We begin with a brief overview of existing work into the dynamics of online communities. We first address these communities in general, before transitioning to a more focused discussion on community evolution.

%\subsection{Social Media and Communities}

Social media communities exist in a variety of forms, and for a variety of purposes. Once established, a community takes on an identity beyond its user base. Unique community characteristics can impact user behavior both within and outside the community. Some communities favor loyalty among users, yielding higher user retention~\cite{hamilton2017loyalty}. Similarly, dynamic and distinctive communities are better at retaining users than communities without these features~\cite{zhang2017community}. On Reddit, subreddits featuring niche content can even prevent users from migrating to another platform~\cite{newell2016user}. However, not all community features are universally predictive; rather, they are unique environments, each governed by a distinct set of rules~\cite{horne2017identifying}. 

% moved this paragraph here from the introduction
Online communities can interact in various interesting ways. In the case of informal communities, topics may shift and overlap as they evolve. In well-defined community spaces, mechanisms exist for inter-community interactions. For example, on Reddit content may be \textit{cross-posted} to multiple subreddits simultaneously, or links to content in another subreddit may fuel additional discussion. Communities may be highly influential, or alternatively dependent, on other communities~\cite{belak2012cross}. Community loyalty can impact not only user behavior within the community, but also the underlying structure of the community itself~\cite{hamilton2017loyalty}. Conversely, how distinctive or dynamic a community is can determine how users engage with that community~\cite{zhang2017community}.  

% stole this paragraph from the intro, then integrated the next paragraph
However, not all community interaction is internal. Online communities must coexist on the same platform as other communities with opposed ideals, which can lead to conflict between communities. Reddit is an archetype of such a Web site where subreddits share, co-mingle, and are frequently in conflict with one another. These disagreements can manifest as conflicting behavior by member users~\cite{datta2019extracting}. Alternatively, users in one subreddit may be mobilized to `attack' another subreddit as a way of expressing negative sentiment, leading to a multi-community conflict~\cite{kumar2018community}.  \nop{Although most community-on-community conflicts are reciprocated, the intensity of the conflict is unlikely to match~\cite{datta2019extracting}.} Content can also be \textit{poached} from small communities and re-posted to a larger community with the goal of amassing Internet karma~\cite{gilbert2013widespread}. Some communities can even become havens of hate speech, prompting bans before the toxic behavior can spread to other communities~\cite{chandrasekharan2017you}. 

% All this is now covered by the above paragraph
% Often these communities exist within the same space, leading to interactions between community entities existing on the same platform. Reddit is an archetype of such a Web site where subreddits share, co-mingle, and are frequently in conflict with one another. These disagreements can manifest as conflicting behavior by member users~\cite{datta2019extracting}, or as more direct invasions~\cite{kumar2018community}. Content can also be \textit{poached} from small communities and re-posted to a larger community with the goal of amassing Internet karma~\cite{gilbert2013widespread}. 

% RIP information diffusion
%In an ecosystem where content is king, information diffuses not only between users, but between communities. For example, most new words are first adopted in large communities before diffusing to smaller ones~\cite{cole2017word}. Unfortunately, most work is focused on diffusion within a single community. Users within the same community have more chances to influence one another in the information diffusion process~\cite{lin2015understanding}, and relevant topics tend to spread from person to person within a single community~\cite{matsumura2003topic}. Diffusion of information from one community to another is a less studied phenomenon.

\newstuff{Within a multi-community ecosystem, patterns of user behavior can predict the types of communities users will choose to engage with. While some users are intrinsically loyal, and are therefore more likely to remain in communities they join initially~\cite{hamilton2017loyalty}, other users may instead wander through communities making them unlikely to stay long-term~\cite{tan2015all}. Models of information forging behavior, for example, rely heavily on a certain stochasticity of the users in order to find alternate paths to additional information~\cite{pirolli2009elementary}. Those who wander risk reaching dead-ends but may also be rewarded by the discovery of new, interesting content. Exploratory, risk-taking browsing appears to be on the rise on the Web. A recent survey of (N=498) Web users found that 12\% of Web browsing was passive, \ie, users had no particular goal in mind, compared to just 2\% in 1997~\cite{liu2020information}. Differences in user information-seeking styles are also predictive: while goal-oriented users engage with a narrower set of communities, wanderers cast a wider net and participate in a greater variety of communities~\cite{waller2019generalists}}%,glenski2017predicting}.}
%In general, users are unlikely to contribute significantly to more than one community, even if they are members of many~\cite{buntain2014identifying}. 
%One user behavior that has not yet been studied is the role of users in creating community connections and links.

\subsection{Evolution of Communities}

Because community-building is a primary motivation for Reddit users to remain on the platform~\cite{moore2017redditors}, users tend to invest their attention in smaller, more specialized communities instead of larger, more general subreddits. To meet this demand, the number of subreddits has expanded to cover increasingly niche topic areas. As a result of this explosion in the number of available communities, the most popular subreddits now account for a smaller percentage of total posts on Reddit~\cite{singer2014evolution}. Overall, active communities have shifted towards smaller, more intimate groups~\cite{ford2021competition}.

As social media has evolved, so too have the communities within it. One aspect of a community's unique identity is the way in which it grows and changes over time. The lifecycle of a community may be characterized with different behavior patterns as the community evolves~\cite{mensah2020characterizing}. Within a single community, users may be more susceptible to community-induced linguistic change during certain community life stages~\cite{danescu2013no}. 

A particular focus of community evolution is the \textit{creation} of these communities -- the gathering of like-minded users into a cohesive group. Existing work in this area is limited, and is generally focused on predicting community popularity, and not on the processes and mechanisms driving community birth and growth. This in itself is a challenging problem, as there are many different ways to measure community success~\cite{cunha2019all}. While many social scientists look to genealogical analysis -- based on users' existing community ties -- to predict future community growth~\cite{tan2018tracing}, others consider both diffusion and other growth to model community growth and death~\cite{kairam2012life}.

It is clear that communities play an important role in how people use and interact with social media, but there remain unanswered questions. How do users find new communities? And how are they created in the first place? The prevailing wisdom is that social media users create new communities to fill gaps in existing offerings, but not all communities will succeed in attracting an active user base. When considering newly-formed communities, how are they established and where do the members of the new subreddit come from? In order for users to self-organize into distinct communities, they must somehow be exposed to new communities as they are created. In this work, we begin to explore the role of subreddit links on Reddit in the community formation and growth process.

\section{Subreddit Links}

On Reddit content is organized into various communities, known as subreddits. Each subreddit hosts various posts, each with associated comments. Within these posts and comments, users may link to content elsewhere on the web, including other Reddit content. One special case of these links is a \textit{subreddit link}, which occurs when a user links directly to a subreddit, instead of a specific post, comment, or external website. These subreddit links can be viewed as a promotion or recommendation of the referenced community. The real example in Fig.~\ref{fig:link_example} illustrates several instances of subreddit links, where each can be identified by the r/ prefix.

In addition to providing recommendation or commentary, subreddit links are also used as commentary or as means for a joke. %\nop{Figure \ref{fig:link_conversations} shows an example of a comment chain consisting solely of subreddit links.} 
A series of links may build upon one another as users add to the joke. Users may also link to a single community as a form of meme, using the subreddit name as a response to the parent post or comment. For example, a link to r/mapswithoutnz is a common occurrence, and serves both as commentary on an inaccurate map, and as promotion of that community. These links may be followed by further references to r/mapswithoutjapan, r/mapswithoutcyprus, or r/mapswithoutthecheasapeakebay -- even though the last of these is not a real subreddit (yet).

% \begin{figure}
%     \centering
%     \includegraphics[width=1.0\linewidth]{temp_figures/link conversation example.png}
%     \caption{Example of a conversation consisting primarily of subreddit links. These links are not only used to promote communities, but are also used as social commentary.}
%     \label{fig:link_conversations}
%     \vspace{.1cm}
% \end{figure}

These subreddit links are not restricted to subreddits that already exist; users may also link to a community that does not exist. Often these types of links are used as jokes, and may follow the naming convention of an existing community. \nop{\todo{example}} This communication method is well-established on Reddit, and users are aware of how these references can create a new subreddit. For example, the community r/birthofasub catalogs new communities that are created following a subreddit link.% In Fig.~\ref{fig:link_example}, r/birthofasub is linked after new subreddit r/simplyprofound is created as a result of the link.

% \todo{here's another example: \nop{https://www.reddit.com/r/tumblr/comments/nyq6ul/adorable/h1mn1kz/} creation of r/CeramicAnimals}

Because of the frequency of these subreddit links, and the ease by which these links can be identified, they are a natural fit for a study on direct community interaction. Each subreddit link is a conscious choice by the posting user to reference a particular subreddit as a whole, instead of a specific post or comment within that community. The unit of information in these subreddit links is therefore the community itself, offering a unique opportunity to assess the impact of these links on community growth and creation.

\subsection{Extracting and Categorizing Links}

\begin{figure}[t]
    \centering
	\includegraphics[]{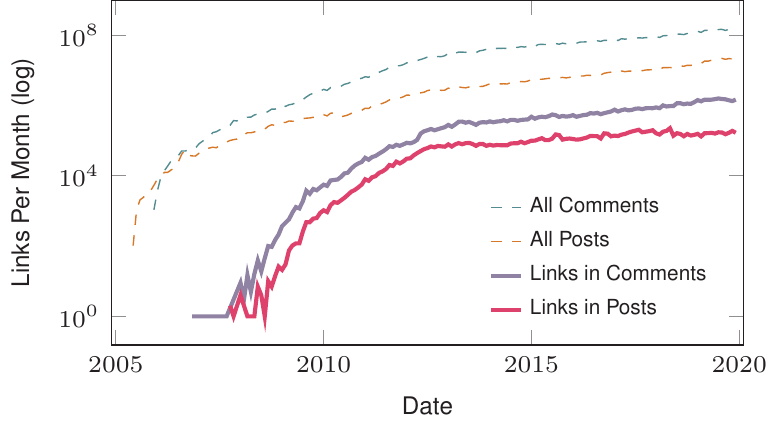}
    \vspace{-.2cm}
    \caption{Number of subreddit links appearing in posts and comments per month. Subreddit links are found in about 1\% of all post and comments after 2010.}
    \label{fig:link_locations}
\end{figure}

We extract all subreddit links on Reddit from June 2005 to December 2019. This includes all links occurring in both posts and comments, by any user, on any subreddit. To be a valid subreddit link, the link must a) reference a subreddit and not a particular post or comment; b) point to a valid subreddit name, according to Reddit's naming requirements\nop{ (as seen in Fig.~\ref{fig:community_name_requirements})}; and c) be properly formatted including the r/ prefix. In this way, we capture all direct community references. In total, we identified 110,365,594 subreddit links.

% \begin{figure}[t]
%     \centering
%     \includegraphics[width=1.0\linewidth]{temp_figures/community name requirements.png}
%     \caption{We only extract links meeting subreddit name requirements, since any link not meeting these criteria could never be a community.}
%     \label{fig:community_name_requirements}
%     \vspace{.1cm}
% \end{figure}

Figure~\ref{fig:link_locations} illustrates the number of subreddit links over time. Although the number of subreddit links per month is only about 1\% of the total comments and posts, their number has grown at about the same rate. We observe that more links occur in comments, rather that in post-text. There are many reasons for this. First, there is a much higher volume of comments than posts on Reddit. Second, many posts do not contain text, and instead consist only of images. Finally, comments tend to be reactive replies to other content, so subreddit links fit more naturally here -- for example, replies to threaded content may spur a reference to a related community.

We categorize these subreddit links into 3 primary types. \textit{Promotional links} are those links that reference a different subreddit that already exists; in this way, the posting user is \textit{promoting} or recommending the referenced community. \textit{Inventive links} are those links to a subreddit that does not exist at the time of the link, but may exist at some point in the future. In this case, the posting user is leveraging the subreddit link format to potentially spawn a new community, or as a referential communication method. Finally, \textit{self-links} are those links from one subreddit to itself. For example, a user may use the link r/AskReddit to refer to the community as a whole within a post in that subreddit. 

\begin{figure}[t]
    \centering
	\includegraphics[]{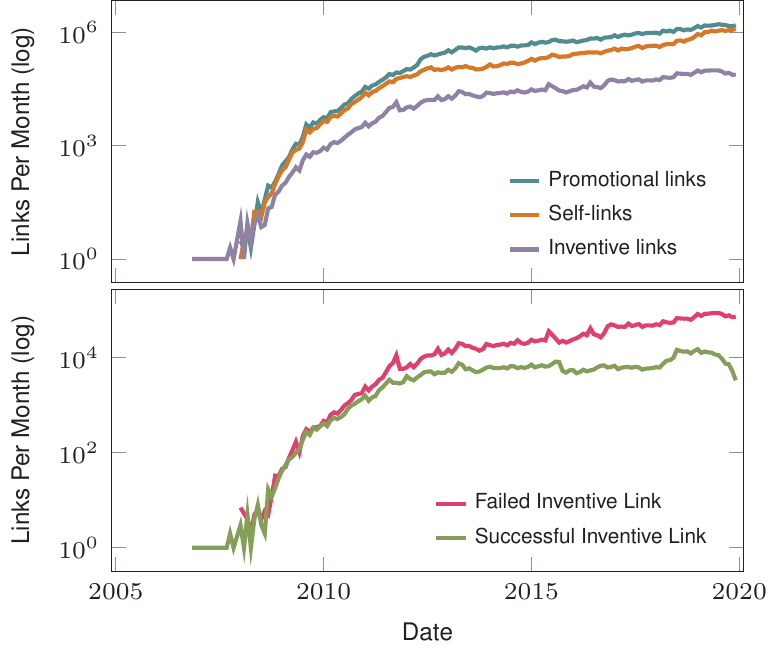}
    \vspace{-.2cm}
    \caption{(Top) Number of subreddit links per month by type. (Bottom) Number of failed and successful inventive links per month.}
    \label{fig:link_counts}
\end{figure}

Figure~\ref{fig:link_counts} shows the frequency of these link types over time. Promotional links are the most common, followed by self-links, and then inventive links. Because self-links do not reference an external community, and therefore do not drive inter-community interactions, we do not analyze these links further.

% \begin{figure}
%     \centering
%     \includegraphics[width=1.0\linewidth]{temp_figures/link_counts_over_time.JPG}
%     \caption{Count of links by type over time. \todo{tex fig} \todo{log scale?} \\Source: link\_counts\_by\_month.xlsx}
%     \label{fig:links_over_time}
%     \vspace{.1cm}
% \end{figure}

The large number of subreddit links and the variety of the different types both indicate the acceptance by users of this interaction method. Users have adopted the link format, using it for the intended purpose -- referencing an existing community -- and co-opting it ironically to link to nonexistent communities sarcastically or as a joke. %Overall, the volume of subreddit links allows for many different kinds of analysis. In the present work, we focus on inventive links and their role in community creation, and on the impact of promotional links on a subreddit's activity levels.

\section{Inventive Links and Community Creation}

The large adoption of this style of communication provides a way to determine the answer to \textbf{RQ1}: Do subreddit links spur the creation of new communities? \newstuff{In the context of the aforementioned models of information seeking behavior, the thinking goes that a hyperlink provides a mode by which the user can move to a new information-space. If a user clicks on a subreddit-link, it shows some modicum of interest in the linked-subject, perhaps framed by the context of the linking information. Then, if it turns out that the community is absent, \ie, the link is broken, then the interested user may be motivated to create the community.}

Formally, we define an \textit{inventive link} as a subreddit link made at time $t^{0}$ in community $u$ referencing a community $v$ that does not exist at time $t^{0}$. In this way, we can say that the posting user may be \textit{inventing} the new subreddit $v$. 

Not all inventive links are successful. Most often, the invented community $v$ is never created as a subreddit. We classify these inventive links as \textit{failed inventive links}. On the contrary, \textit{successful inventive links} are those links that result in the referenced community $v$ being created at some later time $t^{v} > t^{0}$. 

It is important to note that we do not claim to know if the link is directly or causally responsible for the creation of subreddit. Given the observational nature of the present work, we can only study the correlation between links and subreddit creation. Furthermore, we do not impose any time requirements to inventive links; there may exist a long delay between the time of the link $t^{0}$ and the community's creation time $t^v$ (assuming community $v$ is created at all). Therefore, there is no way to know when an inventive link is first made whether it will be successful or not -- only when referenced subreddit $v$ is created can the link be classified as successful. This definition also means that a single community $v$ may experience many inventive links before its creation.

% As with other link types, more inventive links occur in comments than in posts, as shown in Figure \ref{fig:location_inventive}.

% \begin{figure}
%     \centering
%     \includegraphics[width=1.0\linewidth]{temp_figures/inventive_links_by_location.JPG}
%     \caption{Number of inventive links occurring in posts and comments over time. \\Source: link\_counts\_by\_month.xlsx}
%     \label{fig:location_inventive}
%     \vspace{.1cm}
% \end{figure}

\begin{figure}[t]
    \centering
	\includegraphics[]{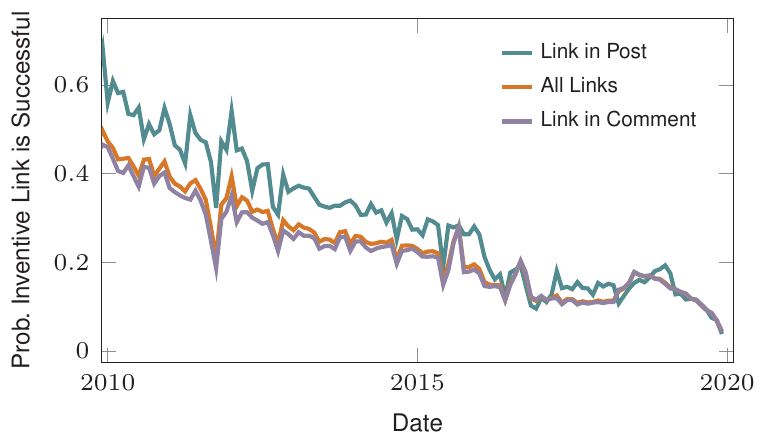}
    \vspace{-.2cm}
    \caption{Probability that a new subreddit is created immediately after an inventive link has decreased over time. This and future plots begin at 2010 because there are too few subreddit links to reliably calculate probabilities before 2010.}
    \label{fig:creation_prob_over_time}
\end{figure}

As indicated in Fig.~\ref{fig:link_counts} and directly illustrated in Fig.~\ref{fig:creation_prob_over_time}, most inventive links are unsuccessful. Even as the overall number of inventive links has increased over time, the number of links resulting in the creation of a new community has remained relatively stable, especially when compared with the growing number of failed inventive links. As a result, the probability a link will produce a new community has decreased over time. Visibility bias may be the reason that inventive links appearing in post text are more likely to result in the creation of a community, because post-text is always displayed at the top of the Web page.

It is important to note our definition of an inventive link is essentially an \textit{attempt} at invention. A single attempt lives permanently -- unless the poster deletes their comment -- but the effect of a single attempt is short-lived. Posts and comments are typically only relevant for a few days after their posting. Subreddit creation statistics bear this out: 8.78\% of all subreddits were created within 1 day of an inventive link; likewise, 10.09\%, 10.62\%, and 11.34\% of all subreddits were created within one, two, and four weeks after an inventive link was posted referencing the not-yet-created new subreddit. If we remove all time restrictions we find that 16.60\% of all created subreddits were referenced by a subreddit link at some point before their creation. \newstuff{Taken together, these statistics suggest that users interpret at least some subreddit links as a signal for a new community to be created. Instead of new subreddits being formed out-of-the-blue, many are spurred by these links. Because a measurable portion of new subreddits experienced inventive links before their creation, this organizational innovation is common enough to merit further investigation.}

\begin{figure}[t]
    \centering
	\includegraphics[]{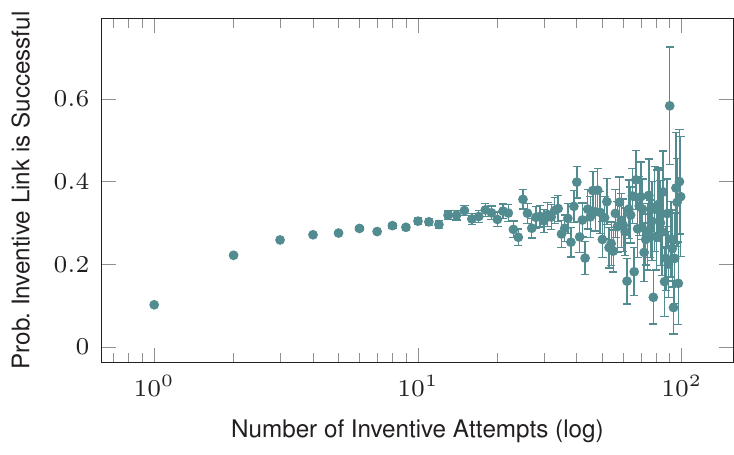}
    \vspace{-.4cm}
    \caption{Probability new subreddit is created given a certain number of inventive links. Error bars represent 95\% confidence from the Binomial test. Repeated links reach a saturation point of 33\% success at about 20 repetitions.}
    \label{fig:creation_prob_by_link_count}
\end{figure}

It is natural to assume that repeated posts of the same inventive link ought to result in a greater likelihood of a successful invention. We therefore ask: do repeated inventive attempts produce an additive effect? To address this question, we plotted the probability that a subreddit is created as a function of the number of inventive attempts in Fig.~\ref{fig:creation_prob_by_link_count}.

There is a logarithmic correlation between the number of inventive attempts and the success probability (Pearson $R^2=0.83$); the probability of a successful invention reaches a saturation point of 33\% at about 20 repetitions. This logarithmic trend follows similar results of saturation in language models and other classical asymptotic plots of diminishing returns~\cite{rowlands2016we}. Overall we found 4,306,117 total inventive links. Of those, 715,106 (13.60\%) were eventually successful and 3,591,011 were not. Note that the existence of duplicate links is responsible for the difference from the 16.60\%-finding from earlier. Here we count subreddit \textit{links}, which contain duplicates; earlier we counted subreddits, which do not.

\subsection{A Natural Experiment for Statistical Testing}

Our definition of a successful inventive link provides that any future creation of a subreddit after some past inventive link is seen as a success -- this is perhaps too permissive. It may therefore be difficult to say whether or not inventive links are correlated with successful subreddit creation from these descriptive statistics alone. Instead, a proper statistical analysis can be realized through a straightforward adoption of survival analysis. 

Formally, given an inventive link to subreddit $v$ at time $t^{0}$, we define survival as a subreddit continuing to \textit{not} exist as time elapses. We define a null hypothesis as follows: for any inventive link to subreddit $v$ at time $t^{0}$ we randomly identify another subreddit $v^\prime$ that was created after $t^{0}$ and define survival as $v^\prime$ not existing as time elapses. \newstuff{Simply put, the null model captures the time to creation of a particular subreddit without an inventive link, as compared to a subreddit experiencing an inventive link.}

\begin{figure}[t]
    \centering
	\includegraphics[]{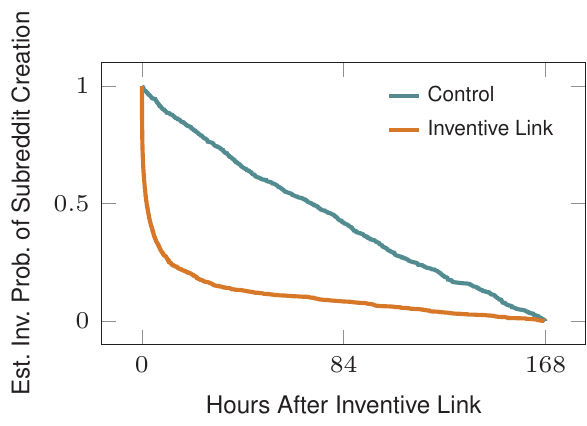}
    \vspace{-.4cm}
    \caption{Survival analysis Kaplan-Meier curve for inventive subreddit links compared to the control. Inventive links significantly increase the probability that the referenced subreddit will be created compared to the control ($\chi^2$=242.61 $p$-value$<$0.001).}
    \label{fig:survival}
\end{figure}

Survival analysis using the log rank test found that inventive links were far more likely to result in the creation of the referenced subreddit within seven days compared to the control ($\chi^2$=242.61 $p$-value$<$0.001). The resulting Kaplan-Meier curve is illustrated in Fig.~\ref{fig:survival}. This figure shows graphically what the statistics report analytically: that an inventive link significantly increases the probability that the referenced subreddit will be created compared to the control. Although not illustrated, the same holds true if we consider time limits of one day ($\chi^2$=971.82 $p$-value$<$0.001) and one month  ($\chi^2$=161.16 $p$-value$<$0.001).

In addition to the log rank test, we also performed Cox regression analysis using the amount of activity in the linking subreddit during the month of the link, as well as the subreddit identities (names) themselves, as covariates. Ultimately, the amount of activity in the linking subreddit was not found to be predictive of the creation of a referenced subreddit. However, subreddits r/nfl, r/AskReddit, and others were found to have slight predictive power on the eventual creation of a referenced subreddit, but that statistical significance waned once Bonferroni correction was applied to the $p$-values. 

Overall, we show that although inventive links are not a guarantee of future subreddit creation, they are highly and immediately correlated to the birth of new communities on Reddit. \newstuff{As information seeking theory suggests, these subreddit links can therefore be seen as a driving force behind organizational innovation: users leverage these links to better shape Reddit's structure and content.}

%\todo{optional more things:
%\begin{itemize}
%    \item how often do inventive links work?
%    \item how many invented communities are actually active? what is lifespan of link-spawned community?
%    \item where do users in new community come from? from subreddit where link originally was?
%    \item why do some inventive links work and some don't?
%%    \item are link-spawned communities related to the linking subreddit?
%    \item how many links (or upvotes?) before community formed?
%    \item look at this paper for methodology clues? %https://ojs.aaai.org/index.php/ICWSM/article/view/14408/14257
%\end{itemize}
%}

\section{Subreddit Links Promote Community Activity}

Having determined that subreddit links are significantly correlated with the creation of new subreddits, our next goal is to answer \textbf{RQ2}: whether or not subreddit links confer significant activity upon the referenced subreddit. \newstuff{Are users successfully driving engagement via community promotion?}

\begin{figure}[t]
    \centering
	\includegraphics[]{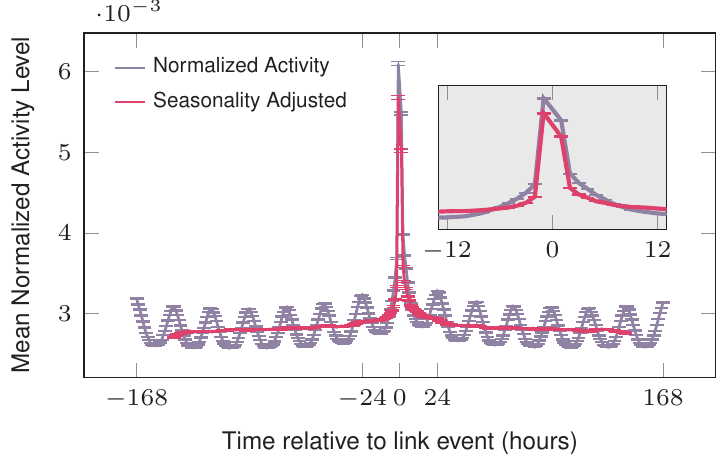}
    \vspace{-.2cm}
    \caption{Subreddit activity levels one week before and after a promotional link in another subreddit (12 hours before and after inset). (Aggregating 10\% of promotional links from 6-2005 through 12-2019.) Error bars show 95\% confidence interval. Activity rises and falls over the course of a 24-hour period, but overall activity is higher following a promotional link event. Removing seasonality from the activity levels makes the rise in activity following a link more apparent.}
    \label{fig:promo_link_activity}
\end{figure}

\begin{figure}[h!]
    \centering
	\includegraphics[]{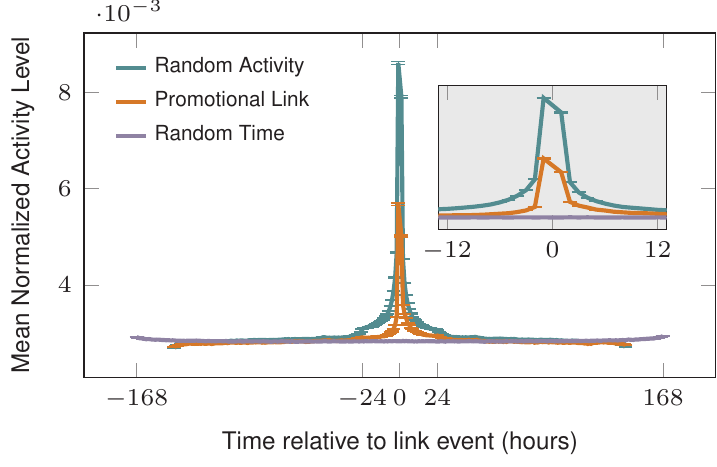}
    \vspace{-.2cm}
    \caption{Activity for one week before and after promotional link (12 hours before and after inset). Error bars show 95\% confidence interval. Activity surrounding a link event is higher than baseline activity around a random time, but lower than activity around a contribution within the subreddit.}
    \label{fig:promo_link_vs_control_diff}
\end{figure}

\begin{figure}[th]
    \centering
	\includegraphics[]{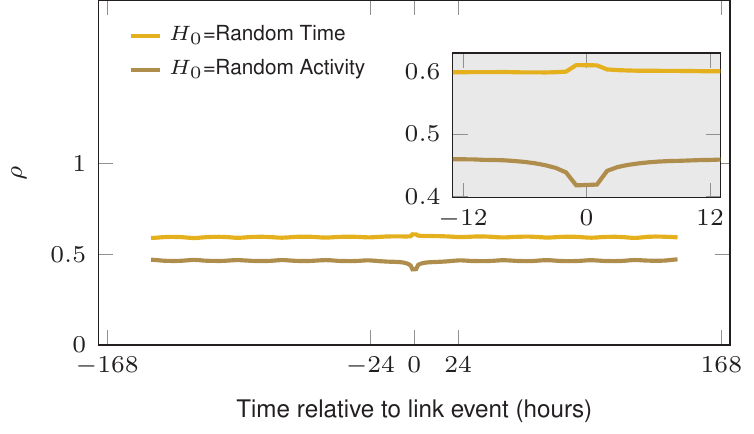}
    \vspace{-.2cm}
    \caption{Mann-Whitney $\rho$ \newstuff{(rho)} statistic between promotional link activity and random controls for one week before and after link (12 hours before and after inset). Extreme values, near 0 or 1, represent stronger distribution separation, while $\rho = 0.5$ indicates complete overlap. We find that a promotion link is correlated with statistically higher activity in the referenced subreddit compared to a random time, but statistically less activity than a random post in the referenced subreddit. \newstuff{Two sided $p<0.001$ for all tests, but are not illustrated here for clarity.} }
    \label{fig:mann-whitney}
\end{figure}

To answer this question we first define a \textit{promotional link} as any subreddit link made at time $t^0$ in community $u$ referencing an \textit{existing} community $v$, where $u \neq v$. By definition, community $v$'s first activity occurred at time $t^v$, where $t^v < t^0$. Therefore, we say that community $u$ is \textit{promoting} community $v$ via a direct link. 

To measure the effect of these promotional links on a subreddit's activity level, we begin by first extracting all promotional link events. Across all of Reddit from June 2005 to December 2019, we found 70,512,949 promotional links. Next, we computed the activity level of the referenced subreddit $v$ surrounding the promotional link. Activity is counted as the number of contributions (posts and comments) in the referenced subreddit, relative to the time of the promotional link. This allows for a comparison of pre-link and post-link activity levels.

% \begin{figure}
%     \centering
%     \includegraphics[width=1.0\linewidth]{temp_figures/2005-06_to_2010-12_normalized_promo_link_activity_7_days_sample_links_0.1_exclude_zeros.png}
%     \caption{Subreddit activity levels before and after a promotional link in another subreddit. (Aggregating 10\% of promo link events from 6-2005 through 12-2010, excluding zero-activity links. \todo{all months!}) Activity rises and falls over the course of a 24-hour period, but overall activity is higher following a promotional link event.}
%     \label{fig:promo_link_activity}
%     \vspace{.1cm}
% \end{figure}

For each promotional link, we normalized the activity levels within each hour for the week before and week after the link posting. \newstuff{Each promotional link is considered independently, even if a subreddit experiences multiple links within a single week.} The normalized activity represents the relative portion of activity that occurs in each hour-bin, thereby controlling for subreddit size effects.

Mean activity levels per hour and their 95\% confidence internals are illustrated in Fig.~\ref{fig:promo_link_activity}. An oscillating activity pattern is clearly evident and represents the daily oscillations in activity dominated by the the primarily English-speaking audience of Reddit, \ie, the Western hemisphere timezones. 

The overall activity level before and after a promotional link is higher than the background. Furthermore, although it is difficult to see from the illustration, the activity level appears to be slightly higher after the link than before, particularly in the first 24 hours. This suggests that promotional links have some effect on subreddit activity levels, but that the effect is not long-lasting or of a high magnitude.

Ultimately, these results illustrate what we already know: activity begets activity~\cite{muchnik2013social,glenski2017rating}. The question is: how much more activity than baseline does a subreddit link provide?

We answer this question by defining two control models: an upper bound and a lower bound. Because activity within a subreddit is known to spur activity within a subreddit, we devise an upper bound null model by drawing a random contribution (post or comment) in subreddit $v$ for each promotional link referencing $v$. As a lower bound null model, we draw one random time in the same month as each subreddit link -- but not responsive to any particular subreddit link or other activity time. Simply put, the upper bound null model corresponds to the activity surrounding a random contribution in the subreddit, and the lower bound corresponds to a random time. We illustrate these normalized and adjusted activity levels in Fig.~\ref{fig:promo_link_vs_control_diff}.

% shiny new multifig!
\begin{figure*}[th!]
    \centering
	\includegraphics[]{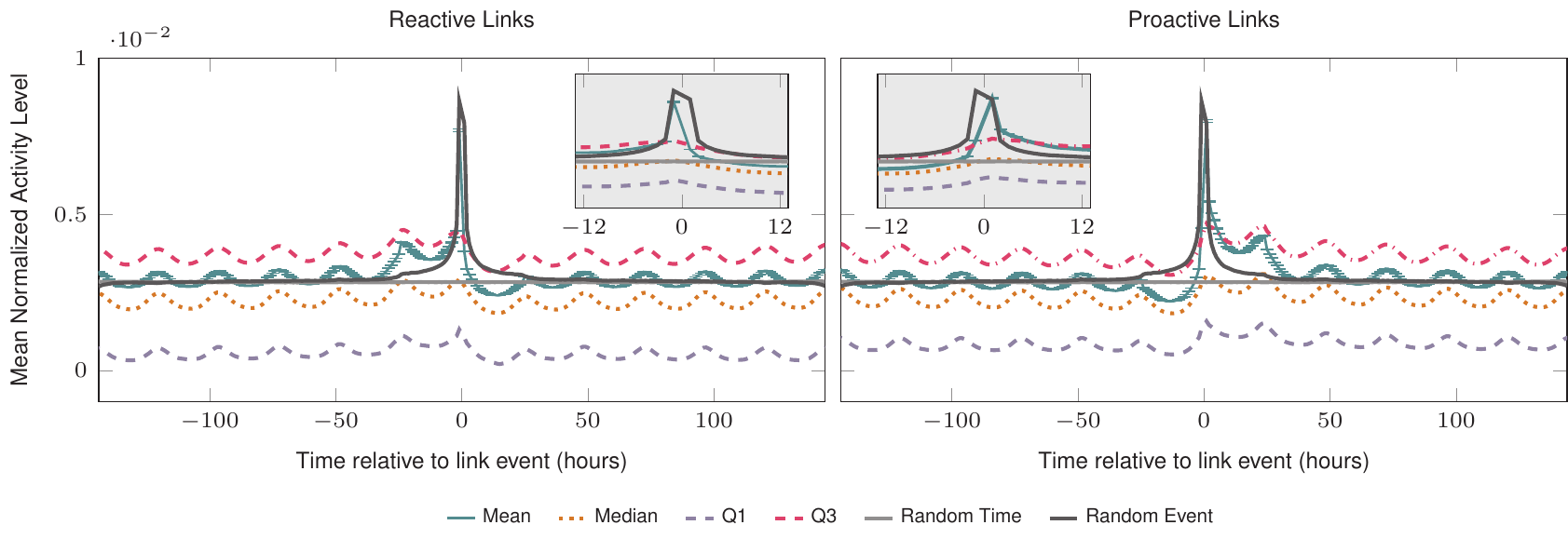}
    \vspace{-.4cm}
    \caption{Average normalized activity level surrounding promotional links partitioned into two categories based on pre-link vs post-link activity levels for one week (12 hours before and after inset). Error bars show 95\% confidence interval. Activity is summed in the 24 hours before and after the promotional link, and each link is categorized based on whether there is more activity before or after the link. In reactive links (left), there is a rise in activity leading up to the promotional link; in proactive links (right), the rise in activity occurs following the link. In both cases, the period of increased activity surpasses the activity of the random contribution control.}
    \label{fig:partition_binary}
\end{figure*}

We find that promotional links generate a large increase in activity in the referenced subreddit. Compared to the random activity upper bound, which appears to be mostly-symmetrical about the selected activity, the Promotional Link activity appears statistically elevated before and after the time of promotional link. %Simply put, promotional links correspond to elevated activity levels in the referenced subreddit.

\newstuff{These findings match predictions from information foraging theory: that a link would drive traffic to the referenced community, just as hyperlinks on the Web drive users to interact with content. However, this behavior is different than a direct content link increasing engagement with an article or video online. Because subreddit links reference an entire community, and not a particular post, the mechanism of engagement is different. For subreddits to exhibit increased activity following a promotional link, users must not only follow the link, but also browse the community before making their own post or comment. This two-step process requires effort beyond following the subreddit link, requiring the user to engage with the community environment first.}

\subsection{Statistical Testing}

Unfortunately, the descriptive statistics presented in the above analysis are unable to provide statistical clarity to RQ2: are promotional links statistically correlated with a change in subreddit activity? To gain further clarity on this question, we apply the Mann-Whitney U-test to the distribution of activity within each hour and compare the activity of promotional links against the upper and lower bounded controls. We plot the Mann-Whitney $\rho$ \newstuff{(rho)} values for the two tests in Fig.~\ref{fig:mann-whitney}. We observe that the activity surrounding random times is significantly lower than promotional links, and that activity surrounding random events is significantly higher than promotional links. \newstuff{All associated two-tailed $p$-values$<$0.001. Simply put, promotional links are correlated with an increase of activity in the referenced subreddit, but not as much as a post-itself.}

\subsection{Types of Promotional Links}

These results raise further questions. Primarily we ask: are promotional links driven by activity in the referenced subreddit? Or do promotional links cause an increase in activity following the link?

\newstuff{We are interested in the driving force behind subreddit links because of what it can tell us about \textit{how} these links are being used. If most promotional links are the result of high activity within the referenced subreddit, this would suggest links are being used to highlight active and thriving communities. This can be seen as users adding structural ties between different types of communities and content, thereby contributing to the organization of Reddit. If, however, most links cause an increase in activity, this would suggest a desire to introduce new users to a different community. These links again create relationships between communities, but also seek to increase engagement in the referenced community. These contrasting motivations represent very different use-cases of promotional links. Absent the ability to survey linking users, our goal is to to better understand these motivations by examining the frequency and activity patterns of different types of links.}

% For this analysis we focus on the 12 hours of activity before a promotional link. For each promotional link, we fit these 12 hours of activity to three different function types: linear, logarithmic, and exponential. Based on which function produces the best fit (based on the smallest total sum squared error), we categorize each link as having linear, logarithmic, or exponential pre-link activity. Then, for each category of link, we compute the average activity level. Results are shown in Figure \ref{fig:partition}. \todo{talk about that figure once we get the other data}

% \begin{figure}
%     \centering
%     \include{plots/promo_partition}
%     \caption{Results of promo links partitioning based on pre-link activity patterns.  \todo{other data} \todo{fix legend} \todo{narrow/zoom?}}
%     \label{fig:partition}
%     \vspace{.1cm}
% \end{figure}

For this analysis we focus on the activity immediately surrounding a promotional link. For each promotional link, we sum the activity in the 24 hours preceding the link, and also the activity in the 24 hours following the link. We then compare these two activity totals. Links with more activity before the link are labeled \textit{reactive} links, and those with more activity after the link are labelled \textit{proactive} links; exact ties are rare and are thrown out from this categorization. \newstuff{Proactive links are most clearly predicted by information foraging theory, wherein a link is used by an information seeker to explore information. However, we also find that many other users leave proverbial breadcrumbs in the form of reactive links to guide others towards interesting activity -- a behavior not predicted by the information foraging framework .}

Across all months of Reddit data, we labeled 4,543,377 links as reactive links, and 3,792,075 links as proactive. \newstuff{We expected to find far more proactive links, \ie, links that motivate activity in the referenced subreddit. However, both link types occur frequently, suggesting two link motivations coexist among Reddit users.}

% old single figure
% \begin{figure}
%     \centering
%     \include{plots/promo_binary_partition}
%     \caption{Average normalized activity level surrounding promotional links partitioned into two categories based on pre-link vs post-link activity levels. Activity is summed in the 24 hours before and after the promotional link, and each link is categorized based on whether there is more activity before or after the link. In activity-driven links, there is a rise in activity leading up to the promotional link; in activity-driving links, the rise in activity occurs following the link. In both cases, the period of increased activity surpasses the activity of the random contribution control. \todo{fix legend} \todo{x-axis range?} \todo{plot quartiles?} \todo{2 graphs, 4 lines each} \todo{smush the outer bits} \todo{plot control}}
%     \label{fig:partition_binary_single}
%     \vspace{.1cm}
% \end{figure}

For both categories of links, we compute the mean, median, Q1, and Q3 of the normalized subreddit activity level. Results are shown in Fig.~\ref{fig:partition_binary}. We find that the reactive vs. proactive categorization is particularly compelling: there is a large discrepancy in activity between these two types of links.
%As expected based on our categorization method, one group has significantly more activity before the link, and the other has more activity after. The two activity curves appear as two sides of the same coin, with activity patterns mirroring one another. For links with more activity before the link, there is a pronounced spike in activity just before the link; for the other category of links, this spike occurs after the link. In addition, there is a temporary change in activity levels in the 12 hours either side of the link event.
%Reactive links could occur when content or conversations in the linked community $v$ encourage direct links in other subreddits. On the other hand, activity-driving links are more likely to be spontaneous, with increased activity as a reaction to new users being exposed to or reminded of the linked community $v$.
%

Does the size of the subreddit matter? For both reactive and proactive links, we examined the distribution of subreddit sizes, for both the linking and referenced subreddits, where subreddit size is defined as the total number of posts and comments per month. Comparing the subreddit size distribution for linking subreddits in the reactive and proactive link groups, we find that these two distributions are very similar (KS-Test $D$ = 0.004, Mann-Whitney $\rho$=0.4979). The distributions of referenced subreddit sizes are also very similar (KS-Test $D$ = 0.023, Mann-Whitney $\rho$=0.5146). 
However, we do find some difference between the subreddit size distributions for linking and referenced subreddit. For reactive links, we find that the linking subreddit tends to be larger than the referenced subreddit (KS-Test $D$ = 0.155, Mann-Whitney $\rho$=0.6184). We find the same relationship for proactive links (KS-Test $D$ = 0.180, Mann-Whitney $\rho$=0.6337). Simply put, subreddit size does not appear to be correlated with a link being reactive or proactive, but linking subreddits tend to be larger than the referenced subreddit.

%cool thanks

Regardless of the motivations and mechanisms behind these two types of links, it is clear that not all promotional links have the same impact on the referenced subreddit. While some links are the cause of increased activity, others appear to be the effect of previous activity. Though linking subreddits tend to be larger than the referenced subreddit, subreddit size is not correlated with the link type. \newstuff{Both types of promotional links allow users to add structural ties between different communities (we assume, however, with different motivations) thereby influencing engagement and content curation beyond simple likes and upvotes.}

\begin{figure*}[th]
    \centering
	\includegraphics[]{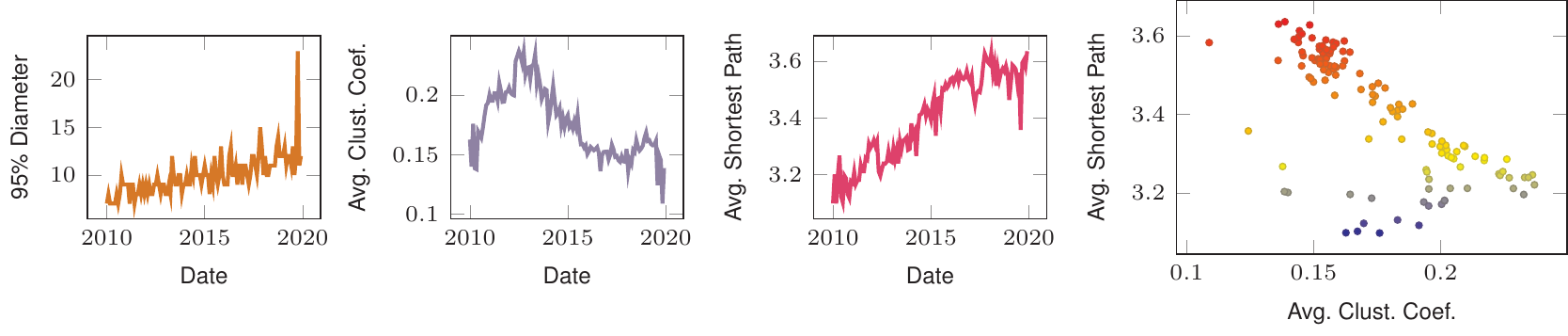}
    \vspace{-.4cm}
    \caption{Subreddit link graph statistics over time. (Left) 95\% effective diameter of the largest weakly-connected component of the subreddit link graph increases slightly over time as new subreddits are created. Because subreddit links have not kept pace with community creation, the diameter of the graph has grown larger. (Left Center) Average clustering coefficient of the largest weakly-connected component increased early in Reddit's life, but then entered a decreasing trend. This suggests that subreddit links are more likely to extend beyond a subreddit's immediate neighborhood. (Right Center) Average shortest path of the largest weakly-connected component increased over time as the graph grew larger as a result of the creation of new subreddits. (Right) Scatter-plot of average clustering coefficient versus average shortest path length. Color indicates the passage of time from blue for 2010 to red for 2019. Over time, the subreddit link graph has become sparser and more treelike.}
    \label{fig:sub_graph_multi}
\end{figure*}

\section{Community Linkage Graph}

Now that we have established the impact of both inventive and promotional subreddit links on the creation of new communities and activity within communities, we turn our attention to the overall community landscape on Reddit. Once again, the many subreddit links provide an avenue and perspective to answer \textbf{RQ3:} How has the landscape of subreddit interactions changed over time? More specifically, we seek to understand the role of subreddit links in modelling Reddit's structure and evolution \newstuff{and to determine the impact of these organically occurring connections.}

% \todo{
% \begin{itemize}
%     \item Graph of subreddit links - all of them ever month by month Directional graph, edge weight = number of links
%     \item look at average distance between communities (clustering coefficient)
%     \item Is it densifying? (95\% diameter) -- Look at “densification in social media”
%     \item subreddit size (activity level) assortativity? something like that? \nop{https://en.wikipedia.org/wiki/Assortativity}
%     % \item clustering coefficient - \nop{https://en.wikipedia.org/wiki/Clustering_coefficient}
%     \item build graph with just the links and subreddits in this month, then new graph next month (do not include subreddits that don't exist, even if they are linked to)
%     % \item maybe someday user assortativity, but that's painful
% \end{itemize}
% }

To assess the changing community landscape on Reddit, we built a \textit{subreddit link graph} based on direct subreddit links. For each month between 2010 and 2019, we considered only inventive and promotional link events occurring within that month. Subreddits with at least one link event (outgoing or incoming) are added to the month's graph as nodes. A directed edge is added to the graph for each subreddit link, from the linking subreddit to the referenced subreddit. Edges are weighted with the number of links to manage repeated links between the same pair of subreddits. We intentionally chose to build monthly versions of the subreddit graph rather than a single cumulative graph. The monthly subreddit graphs provide temporal snapshots into the dynamics and evolution of the macro-network. A cumulative graph (combining all previous months together) would potentially mix relationships that occurred years apart and are no longer relevant. 

We computed a variety of graph statistics on the monthly subreddit link graphs to evaluate how the connectivity patterns of Reddit have evolved. This macro-view provides insights into how communities, and the ties between them, have changed over time.

The four subplots in Fig.~\ref{fig:sub_graph_multi} illustrate these dynamics. The subreddit graph consisted of a giant weakly connected component that comprised about 98\% of all nodes (but varied slightly ($\pm$1\%) month to month). The diameter plot in Fig.~\ref{fig:sub_graph_multi} (left) shows how the 95\% effective diameter of the giant component has changed overtime. Generally, the diameter has increased slightly overall. This is in contrast to the well-established shrinking diameter's of person-to-person social networks~\cite{leskovec2007graph}. These contrasting results are likely due to the creation of new subreddits: as more communities are created and participate in subreddit links, the largest weakly-connected component appears to be growing. 

We also computed the density and subreddit activity assortativity for these monthly graphs, but do not illustrate them here. Both density and assortativity of the graphs were remarkably stable. Specifically, the density of the graph was quite low: ranging from 0.1--0.01\%. Graph assortativity was computed using the total activity (number of posts and comments) of each subreddit in each month. Assortativity measures the preference for a network's nodes to connect to other similar nodes; assortativity near 1 indicates perfect assortativity, \ie, active nodes link to other active nodes; values near -1 indicate a disassortative network, \ie, active nodes link to inactive nodes. Overall, we found that the activity assortativity was near 0 and relatively stable over all months.

Independently, the average clustering coefficient graph (center left) and the average shortest path length (center right) shows that the subreddit graph tends to cluster together less over time. However, much information can be gleaned by consider considering these two statistics in tandem~\cite{you2020graph} as shown in Fig.~\ref{fig:sub_graph_multi} (right). This figure plots points from blue to red corresponding to 2010 to 2019 (more red is more recent). We interpret the red-shift in this graph to indicate that the subreddit graph is becoming sparser and more treelike over time. This suggests that subreddit links are not keeping pace with the creation of new subreddits, and that community ties do not cluster as strongly they did in the past.

% The density plot (on left) shows that the density of the Reddit has decreased significantly over time. Even with increasing numbers of subreddit links, the ever-growing number of communities on Reddit means that the overall graph density has decreased. This suggests that communities on Reddit are less tightly connected or related than they were in the past. 

%The weakening relational structure of Reddit may be due to subreddits becoming more specialized~\cite{ford2021competition} or simply because subreddit growth has far outpaced subreddit link growth..

\newstuff{Because Reddit is a user-driven social network, all content is created, curated, and organized by the participating users. New subreddits can be created with relatively little effort, leading to subreddits becoming more specialized over time~\cite{ford2021competition, singer2014evolution}. This expansion of Reddit mirrors the behavior of individual users, who continually post in new communities as they wander for new content~\cite{tan2015all} -- new subreddits are required to feed these wandering users. \nop{However, this expansion contrasts with some users' tendencies to remain with familiar communities instead of migrating to new ones~\cite{bergstrom2021reddit}. \todo{why cite no work?}} As the organizational structure of Reddit has grown to encompass more different types of content, and Reddit has evolved and shifted focus~\cite{singer2014evolution}, subreddit links serve as an another way for users to innovate and shape the environment to their liking by influencing the community landscape directly.}

%What do these links tell us about the changing role of communities on Reddit? 

\nop{\begin{figure}
    \centering
    \include{plots/sub_graph_density}
    \caption{Density of the subreddit link graph decreases over time, indicating that subreddit links do not keep up with the creation of new subreddits. \todo{x-axis labels, no commas}}
    \label{fig:link_graph_density}
    \vspace{.1cm}
\end{figure}}

\nop{
\begin{figure}
    \centering
    \include{plots/sub_graph_diameter}
    \caption{Diameter of the largest weakly-connected component of the subreddit link graph increases slightly over time as new subreddits are created. Because subreddit links do not keep pace with community creation, the diamter of the graph necessarily grows larger. \todo{x-axis labels, no commas}}
    \label{fig:link_graph_diameter}
    \vspace{.1cm}
\end{figure}
}

\nop{
\begin{figure}
    \centering
    \include{plots/sub_graph_assort}
    \caption{Assortativity on subreddit size is negative, but increases over time, approaching a ceiling at 0, indicating the subreddit link graph has become less assortative. \todo{x-axis labels, no commas}}
    \label{fig:link_graph_assort}
    \vspace{.1cm}
\end{figure}
}

\section{Conclusions}

% \todo{
% \begin{itemize}
%     \item links exist, people use them
%     \item used for promotion, invention, jokes, memes, commentary, etc
%     \item impact of links on community landscape as a whole (subreddit graph)
%     \item some links spawn new communities
%     \item other links change the character/activity in a subreddit
%     \item links have a life of their own
% \end{itemize}
% }

On Reddit, individual subreddits are used to organize content and connect users. Because these communities exist within the same space, and must therefore compete for attention, content, and users, subreddits often interact with one another. One mode of interaction is the \textit{subreddit link}, which occurs when a user makes a direct reference to a subreddit in another community. \newstuff{Information foraging theory suggests that these links can be seen as promotion of the referenced community, driving information seekers to that content. But we also find that subreddit links have evolved into a mechanism for jokes, memes, or other commentary. Subreddit links can also be seen as a new form of innovation, granting users the power to influence community engagement and the organization of content.} Based on the frequency and ubiquity of these links, we have undertaken a study on subreddit links on Reddit, with the goal of understanding their impact on both the referenced subreddit, and on the subreddit landscape as a whole.

Subreddit links may reference communities that do not yet exist. These \textit{inventive links} can be seen as users pointing out a gap in existing subreddit offerings. In some cases, new subreddits are created as a result of these links, though not all subreddits originate in this way. %\nop{Although repeated inventive links are possible, and theoretically increase the visibility of the suggested community, we find that repeated links to the same subreddit do not increase the probability that the community will be created. Overall, the probability of an inventive link spawning a new community has decreased over time. This suggests that as Reddit has matured, the need for new subreddits has decreased.}
We find that these inventive links are significant drivers of new subreddits and that inventive links have a major impact the overall community ecosystem of Reddit. 

In contrast, \textit{promotional links} reference a subreddit that already exists; these links can be seen as the posting user promoting or recommending the referenced community \newstuff{in an attempt to influence engagement}. We find that promotional links induce a temporary increase in activity levels surrounding the link, particularly in the first few hours after the link. More specifically, activity levels around a promotional link are higher than baseline subreddit activity levels. Additionally, we find that some links are \textit{reactive} -- driven by other activity, while others are \textit{proactive} -- driving activity in the referenced subreddit. This suggests that there are different motivations for links depending on the subreddit and situation. Surprisingly, subreddit size does not impact whether a link is reactive or proactive, but linking subreddits do tend to be more active than the referenced subreddit. %\newstuff{These findings confirm that subreddit links allow users to influence content on Reddit beyond the more typical score-based curation.}

Finally, we examined the changing community landscape on Reddit via a \textit{subreddit link graph}, constructed based on subreddit links. We find that this graph has become more sparse and less clustered over time. This indicates that as Reddit has grown and new subreddits were created, the number and variety of subreddit links has not kept pace, leading to an effective decrease in the reach of subreddit links. %\newstuff{Subreddit links may be viewed as users' attempts to impose their own desires on the organization of communities online.}

% \todo{overall conclusions here - what's the big takeaway? something about links as the mechanism for community interaction and communication within a changing social media landscape}

\newstuff{In their totality, these results confirm and expand upon existing theories of information seeking behavior: subreddit links on Reddit not only serve as links to information, but have measurable impact on the creation of new communities and the activity within existing subreddits. By serving as a form of innovation, these links allow users to influence community organization and engagement. Users have adopted the subreddit link, and leverage it for a variety of purposes, shaping the landscape of Reddit as a result. However, some results also stand in contrast to the well-established social network dynamics that predict shrinking diameters of social networks. Instead, we find that the subreddit network appears to be expanding, not shrinking. }

\section{Future Work}

We are convinced that subreddit links are a powerful tool for the study of community interaction, but recognize the limitations of the presented work. A similar analysis could be applied to other social media platforms, like Facebook, Twitter, and Instagram, but significant modification would be required as not all platforms have such well-defined communities. \nop{For example, although Facebook has groups, Twitter does not have a community mechanism as part of the platform.} \newstuff{There are also other mechanisms of community promotion that could be studied and compared. For example, crossposts on Reddit allow the same content to be shared to many different communities, which can be seen as a less direct method of community promotion.}

\nop{Further research into community interactions is also merited.} 
Though we have shown that subreddit links impact the activity levels within the referenced community, further investigation is required to determine the true cause and source of the additional activity. Are new users joining the community the cause of elevated activity levels? Were these users previously members of the linking community? Or is the increased activity in the referenced subreddit driven by existing community members? Answering these questions would help us more-thoroughly understand the power of community links.

We are also interested in the types of users creating these community links. Are linking users active in both the linking and referenced community? Are linking users long-established members of the linking community, or did they join only to recommend to another community? These findings could assist creators of new communities in growing their following. \newstuff{Similarly, we are interested in what \textit{types} of contexts attract subreddit links. Are there particular discussion dynamics that are more likely to produce a subreddit link? Do the links themselves change the pattern of discussion? Understanding the complex interactions between users, communities, and discussions would allow social media developers to better tailor the user experience to individual needs and preferences.}

%\done{While answering RQ1, it would be interesting to understand better when an inventive link is successful if covariates can predict it. For example, is the advertising community's topic (or size) predictive of a new community creation?}
    
%\done{Another direction could be to investigate how a community created after an inventive link compares to a random community created around the same time in Reddit? How do they compare in terms of size, activity levels, etc.? Does this new community created after an inventive link have any initial advantage because it had an audience in the post/comment it was mentioned?}

Finally, we would like to tease out the factors that make some subreddit links more successful than others. We find that repeated links are correlated with a greater likelihood of a subreddit's creation, but, do repeated links guarantee community growth? \newstuff{Does the size of the linking subreddit determine link visibility, and therefore impact?} Are links more beneficial when the referenced community is small, or when it is large? \newstuff{Are upvotes on a link, or comment author popularity, predictive factors? Are communities created following an inventive link more or less likely to succeed?} There are a large variety of factors that remain unexplored in the understanding of how and why subreddit communities grow and evolve, which we plan to explore in future studies.

% \todo{
% \begin{itemize}
%     \item more platforms - facebook, twitter, instagram?
%     \item long term effect of repeated links
%     \item user-level analysis - do all users link? few? are linking users active in both communities?
%     \item further research into community interactions
%     \item quantify change in activity level? how big? how long?
%     \item which users are 'following' these links, and becoming active in the promoted community? were they previously active in the promoting community?
%     \item how does subreddit size affect this - both of the promoting subreddit, and the promoted subreddit?
%     \item link partition sizes over time? is there a shift towards a single `type' of link?
% \end{itemize}
% }

\section{Broader Impacts and Ethical Considerations}

%\done{In order to provide a balanced perspective, authors are required to include a statement about the potential broader impact of their work and ethical considerations. This statement needs to, at the bare minimum, be presented in a clearly marked paragraph/subsection/section in your paper. Authors should take care to discuss both positive and negative outcomes. Authors are also expected to describe steps taken to prevent or mitigate potential negative outcomes. For full papers that have collected new datasets and for dataset papers, discuss ethical considerations in the data collection process, and considerations around its release, and both potentially positive and negative outcomes of its use by others.}

% checklist questions come from NeurIPS 2021 Paper Checklist Guidelines
% https://neurips.cc/Conferences/2021/PaperInformation/PaperChecklist

% \todo{Do the main claims made in the abstract and introduction accurately reflect the paper's contributions and scope?}

\newstuff{
The results presented in this paper show that subreddit links have a substantive impact on community creation and activity levels within established communities on Reddit. Based on these findings, we recognize the possible applications, both good and bad. We therefore outline here the potential impacts and strategies to mitigate them. \nop{Moving forward, we hope to more specifically quantify the predictive power of these subreddit links, and to investigate similar phenomena on other platforms.}}

\subsection{Potential Impacts}

% \todo{Did you discuss any potential negative societal impacts of your work?}
% \todo{Did you discuss any potential positive societal impacts of your work? (I added this one based on ICWSM requirements.)}

\newstuff{
As with most social media research, the work presented here is subject to unique ethical considerations. The line between positive and negative outcomes is often very thin in this domain, and the potential uses of subreddit links are no different. Intention and nuance are critical in evaluating the ethics of these community promotions, as the mechanism of use is the same for helpful and harmful instances.}

\newstuff{
Given that we have shown how subreddit links influence the community landscape on Reddit, one possible negative outcome is the manipulation of these subreddit links by bad actors to push a particular agenda or viewpoint. Community-focused misinformation contributes to siloing and echo chambers, fueling conflict. Similarly, users could co-opt common links by creating a community whose purpose does not match the subreddit name, potentially exposing other users to harmful or malicious content. To mitigate these negative impacts, Reddit and other social media platforms may need to monitor the frequency of subreddit links and the formation of new communities to detect and intervene in cases of misuse.}

\newstuff{
There are also positive outcomes that could result from this work. First, a better understanding of the power of subreddit links could allow for new avenues of content promotion. For example, users can direct other commenters to specific helpful subreddits based on the context of the discussion. This is already a common use case of subreddit links, but with concrete evidence of link impact, organizations could adopt the convention for good, like recommendation of mental health communities. Subreddit links could also be leveraged in the fight against misinformation and disinformation, by directing users to communities with more rigorous moderation and fact-checking.}

\subsection{Limitations}

% \todo{Did you describe the limitations of your work?}

\newstuff{These findings are limited to subreddit communities on Reddit. These findings are likely to generalize to other social media platforms, but further experiments are required to confirm. Our experiments assume that all subreddit links are deliberate choices made by users, and therefore may incorrectly capture some links that were mistakes or typos. However, even if a subreddit link was made by mistake, any well-formed link will point to a subreddit that may or may not exist. In compiling the link dataset, we considered all well-formed links, without filtering for link score or visibility, community size, or other factors that could impact a link's effectiveness. Future work investigating these nuances may reveal trends or effect not discovered in this aggregate analysis.}

% \todo{If you are including theoretical results: Did you state the full set of assumptions of all theoretical results? Did you include complete proofs of all theoretical results?}

\nop{\newstuff{No theoretical results.}}

\subsection{Data Collection and Privacy Considerations}

% \todo{If you ran experiments:  (a) Did you include the code, data, and instructions needed to reproduce the main experimental results (either in the supplemental material or as a URL)? (b) Did you specify all the training details (e.g., data splits, hyperparameters, how they were chosen)? (c) Did you report error bars (e.g., with respect to the random seed after running experiments multiple times)? (d) Did you include the amount of compute and the type of resources used (e.g., type of GPUs, internal cluster, or cloud provider)?}

\newstuff{Because this data contains artifacts of human subjects, we sought human subjects review from the ethical review board at the University of Notre Dame; an exemption was granted for this study on the basis that all data was public and no treatments were conducted by the research team. Additionally, most Reddit users act anonymously, but any personal information was willingly shared by the user. All results presented are in aggregate. We did not seek nor attempt to de-anonymize any data and therefore do not reveal any private or personally identifying information.}

\newstuff{The dataset used for experiments is available from \url{files.pushshift.io/reddit}. We encourage interested parties to contact us for more details on dataset compilation and analysis. \nop{No machine learning models were used, so hyperparameters are other training artifacts not applicable.} All plots include confidence intervals as appropriate. All experiments were conducted on a local high-capacity CPU machine.}


\begin{thebibliography}{33}
\providecommand{\natexlab}[1]{#1}
\providecommand{\url}[1]{\texttt{#1}}
\providecommand{\urlprefix}{URL }
\expandafter\ifx\csname urlstyle\endcsname\relax
  \providecommand{\doi}[1]{doi:\discretionary{}{}{}#1}\else
  \providecommand{\doi}{doi:\discretionary{}{}{}\begingroup
  \urlstyle{rm}\Url}\fi

\bibitem[{Bairi, Carman, and Ramakrishnan(2015)}]{bairi2015evolution}
Bairi, R.; Carman, M.; and Ramakrishnan, G. 2015.
\newblock On the evolution of Wikipedia: dynamics of categories and articles.
\newblock In \emph{ICWSM}, volume~9.

\bibitem[{Bel{\'a}k, Lam, and Hayes(2012)}]{belak2012cross}
Bel{\'a}k, V.; Lam, S.; and Hayes, C. 2012.
\newblock Cross-community influence in discussion fora.
\newblock In \emph{ICWSM}, volume~6.

\bibitem[{Chandrasekharan et~al.(2017)Chandrasekharan, Pavalanathan,
  Srinivasan, Glynn, Eisenstein, and Gilbert}]{chandrasekharan2017you}
Chandrasekharan, E.; Pavalanathan, U.; Srinivasan, A.; Glynn, A.; Eisenstein,
  J.; and Gilbert, E. 2017.
\newblock You can't stay here: The efficacy of reddit's 2015 ban examined
  through hate speech.
\newblock \emph{CSCW} 1:
  1--22.

\bibitem[{Cunha et~al.(2019)Cunha, Jurgens, Tan, and Romero}]{cunha2019all}
Cunha, T.; Jurgens, D.; Tan, C.; and Romero, D. 2019.
\newblock Are all successful communities alike? Characterizing and predicting
  the success of online communities.
\newblock In \emph{TheWebConference}, 318--328.

\bibitem[{Danescu-Niculescu-Mizil et~al.(2013)Danescu-Niculescu-Mizil, West,
  Jurafsky, Leskovec, and Potts}]{danescu2013no}
Danescu-Niculescu-Mizil, C.; West, R.; Jurafsky, D.; Leskovec, J.; and Potts,
  C. 2013.
\newblock No country for old members: User lifecycle and linguistic change in
  online communities.
\newblock In \emph{TheWebConference}, 307--318. ACM.

\bibitem[{Datta and Adar(2019)}]{datta2019extracting}
Datta, S.; and Adar, E. 2019.
\newblock Extracting inter-community conflicts in reddit.
\newblock In \emph{ICWSM}, volume~13, 146--157.

\bibitem[{Farzan and Brusilovsky(2018)}]{farzan2018social}
Farzan, R.; and Brusilovsky, P. 2018.
\newblock Social navigation.
\newblock \emph{Social information access} 142--180.

\bibitem[{Fisher, Edelez, and McKechnie(2005)}]{fisher2005theories}
Fisher, KE.; Erdelez, S.; and McKechnie, LEF. 2005.
\newblock \emph{Theories of information behavior} Information Today, Inc.

\bibitem[{Ford, Krohn, and Weninger(2021)}]{ford2021competition}
Ford, T.; Krohn, R.; and Weninger, T. 2021.
\newblock Competition Dynamics in the Meme Ecosystem.
\newblock \url{https://arxiv.org/abs/2102.03952}.

\bibitem[{Gao et~al.(2021)Gao, Cui, Bulut, Zhai, and Chen}]{gao2021examining}
Gao, Y.; Cui, Y.; Bulut, O.; Zhai, X.; and Chen, F. 2021.
\newblock Examining adults’ web navigation patterns in multi-layered
  hypertext environments.
\newblock \emph{Computers in Human Behavior} 107142.

\bibitem[{Gilbert(2013)}]{gilbert2013widespread}
Gilbert, E. 2013.
\newblock Widespread underprovision on reddit.
\newblock In \emph{CSCW}, 803--808.

\bibitem[{Glenski and Weninger(2017)}]{glenski2017rating}
Glenski, M.; and Weninger, T. 2017.
\newblock Rating effects on social news posts and comments.
\newblock \emph{ACM Trans. on Intelligent Systems and Technology} 8(6): 78.

\bibitem[{Hamilton et~al.(2017)Hamilton, Zhang, Danescu-Niculescu-Mizil,
  Jurafsky, and Leskovec}]{hamilton2017loyalty}
Hamilton, W.; Zhang, J.; Danescu-Niculescu-Mizil, C.; Jurafsky, D.; and
  Leskovec, J. 2017.
\newblock Loyalty in online communities.
\newblock In \emph{ICWSM}, volume~11.

\bibitem[{Horne, Adali, and Sikdar(2017)}]{horne2017identifying}
Horne, B.~D.; Adali, S.; and Sikdar, S. 2017.
\newblock Identifying the social signals that drive online discussions: A case
  study of reddit communities.
\newblock In \emph{ICCCN}, 1--9. IEEE.

\bibitem[{Jackson, Bailey, and Foucault~Welles(2018)}]{jackson2018girlslikeus}
Jackson, S.~J.; Bailey, M.; and Foucault~Welles, B. 2018.
\newblock \# GirlsLikeUs: Trans advocacy and community building online.
\newblock \emph{New Media \& Society} 20(5): 1868--1888.

\bibitem[{Kairam, Wang, and Leskovec(2012)}]{kairam2012life}
Kairam, S.~R.; Wang, D.~J.; and Leskovec, J. 2012.
\newblock The life and death of online groups: Predicting group growth and
  longevity.
\newblock In \emph{WSDM}, 673--682.

\bibitem[{Kumar et~al.(2018)Kumar, Hamilton, Leskovec, and
  Jurafsky}]{kumar2018community}
Kumar, S.; Hamilton, W.~L.; Leskovec, J.; and Jurafsky, D. 2018.
\newblock Community interaction and conflict on the web.
\newblock In \emph{TheWebConference},
  933--943.

\bibitem[{Leskovec, Kleinberg, and Faloutsos(2007)}]{leskovec2007graph}
Leskovec, J.; Kleinberg, J.; and Faloutsos, C. 2007.
\newblock Graph evolution: Densification and shrinking diameters.
\newblock \emph{ACM Transactions on Knowledge Discovery from Data (TKDD)} 1(1):
  2--es.

\bibitem[{Liu(2020)}]{liu2020information}
Liu, F. 2020.
\newblock How Information-Seeking Behavior Has Changed in 22 Years.
\newblock
  \urlprefix\url{\url{https://www.nngroup.com/articles/information-seeking-behavior-changes/}}.
\newblock [Online; posted 20-January-2020].

\bibitem[{Mensah, Xiao, and Soundarajan(2020)}]{mensah2020characterizing}
Mensah, H.; Xiao, L.; and Soundarajan, S. 2020.
\newblock Characterizing the Evolution of Communities on Reddit.
\newblock In \emph{International Conference on Social Media and Society},
  58--64.

\bibitem[{Moore and Chuang(2017)}]{moore2017redditors}
Moore, C.; and Chuang, L. 2017.
\newblock Redditors revealed: Motivational factors of the Reddit community.
\newblock In \emph{HICSS}.

\bibitem[{Muchnik, Aral, and Taylor(2013)}]{muchnik2013social}
Muchnik, L.; Aral, S.; and Taylor, S.~J. 2013.
\newblock Social influence bias: A randomized experiment.
\newblock \emph{Science} 341(6146): 647--651.

\bibitem[{Newell et~al.(2016)Newell, Jurgens, Saleem, Vala, Sassine, Armstrong,
  and Ruths}]{newell2016user}
Newell, E.; Jurgens, D.; Saleem, H.; Vala, H.; Sassine, J.; Armstrong, C.; and
  Ruths, D. 2016.
\newblock User migration in online social networks: A case study on reddit
  during a period of community unrest.
\newblock In \emph{ICWSM}, volume~10.

\bibitem[{Pirolli(2009)}]{pirolli2009elementary}
Pirolli, P. 2009.
\newblock An elementary social information foraging model.
\newblock In \emph{CHI}, 605--614.

\bibitem[{Rowlands, Waddell, and McKenna(2016)}]{rowlands2016we}
Rowlands, T.; Waddell, N.; and McKenna, B. 2016.
\newblock Are we there yet? A technique to determine theoretical saturation.
\newblock \emph{Journal of Computer Information Systems} 56(1): 40--47.

\bibitem[{Singer et~al.(2014)Singer, Fl{\"o}ck, Meinhart, Zeitfogel, and
  Strohmaier}]{singer2014evolution}
Singer, P.; Fl{\"o}ck, F.; Meinhart, C.; Zeitfogel, E.; and Strohmaier, M.
  2014.
\newblock Evolution of reddit: from the front page of the internet to a
  self-referential community?
\newblock In \emph{TheWebConference}, 517--522.

\bibitem[{Tan(2018)}]{tan2018tracing}
Tan, C. 2018.
\newblock Tracing community genealogy: how new communities emerge from the old.
\newblock In \emph{ICWSM}, volume~12.

\bibitem[{Tan and Lee(2015)}]{tan2015all}
Tan, C.; and Lee, L. 2015.
\newblock All who wander: On the prevalence and characteristics of
  multi-community engagement.
\newblock In \emph{TheWebConference}, 1056--1066.

\bibitem[{Waller and Anderson(2019)}]{waller2019generalists}
Waller, I.; and Anderson, A. 2019.
\newblock Generalists and specialists: Using community embeddings to quantify
  activity diversity in online platforms.
\newblock In \emph{TheWebConference}, 1954--1964.

\bibitem[{Weninger, Johnston, and Han(2013)}]{weninger2013parallel}
Weninger, T.; Johnston, T.~J.; and Han, J. 2013.
\newblock The parallel path framework for entity discovery on the web.
\newblock \emph{ACM Transactions on the Web (TWEB)} 7(3): 1--29.

\bibitem[{West and Leskovec(2012)}]{west2012human}
West, R.; and Leskovec, J. 2012.
\newblock Human wayfinding in information networks.
\newblock In \emph{TheWebConference}, 619--628.

\bibitem[{Yang et~al.(2019)Yang, Kraut, Smith, Mayfield, and
  Jurafsky}]{yang2019seekers}
Yang, D.; Kraut, R.~E.; Smith, T.; Mayfield, E.; and Jurafsky, D. 2019.
\newblock Seekers, providers, welcomers, and storytellers: Modeling social
  roles in online health communities.
\newblock In \emph{CHI}, 1--14.

\bibitem[{You et~al.(2020)You, Leskovec, He, and Xie}]{you2020graph}
You, J.; Leskovec, J.; He, K.; and Xie, S. 2020.
\newblock Graph structure of neural networks.
\newblock In \emph{ICML}, 10881--10891.
  PMLR.

\bibitem[{Zhang et~al.(2017)Zhang, Hamilton, Danescu-Niculescu-Mizil, Jurafsky,
  and Leskovec}]{zhang2017community}
Zhang, J.; Hamilton, W.; Danescu-Niculescu-Mizil, C.; Jurafsky, D.; and
  Leskovec, J. 2017.
\newblock Community identity and user engagement in a multi-community
  landscape.
\newblock In \emph{ICWSM}, volume~11.

\end{thebibliography}
\end{document}